\documentclass{jetpl}

\usepackage[nospace]{cite}

\twocolumn

\lat

\title{Magnetization of the Mn$_{1-x}$Fe$_{x}$Si in high magnetic field up to 50~T: possible evidence of a field-induced Griffiths phase}

\rtitle{Magnetization of the Mn$_{1-x}$Fe$_{x}$Si in high magnetic field up to 50~T: ...}

\sodtitle{Magnetization of the Mn$_{1-x}$Fe$_{x}$Si in high magnetic field up to 50~T: possible evidence of a field-induced Griffiths phase}

\author{
	S.V.~Demishev $^{1,2}$,
	A.N.~Samarin $^{1,3}$,
	J.~Huang $^{4}$,
	V.V.~Glushkov $^{1,3}$,
	I.I.~Lobanova $^{1,3}$,
	N.E.~Sluchanko $^{1,3}$,
	N.M.~Chubova $^{5}$,
	V.A.~Dyadkin $^{5,6}$,
	S.V.~Grigoriev $^{5,7}$,
	M.Yu.~Kagan $^{2,8}$,
	J.~Vanacken $^{4}$,
	and V.V.~Moshchalkov $^{4}$
}

\rauthor{S.V.~Demishev, A.N.~Samarin, J.~Huang}

\sodauthor{Demishev, Samarin, Huang}

\address{
	$^1$Prokhorov General Physics Institute of RAS, Vavilov street, 38, 119991 Moscow, Russia\\~\\
	$^2$National Research University Higher School of Economics, Myasnitskaya street, 20, 101000 Moscow, Russia\\~\\
	$^3$Moscow Institute of Physics and Technology, 9 Institutsky lane, Dolgoprudny 141700 Moscow region, Russia\\~\\
	$^4$KU Leuven, Department of Physics and Astronomy, Celestijnenlaan 200D, B-3001 Leuven, Belgium\\~\\
	$^5$Petersburg Nuclear Physics Institute, Gatchina, 188300 Saint-Petersburg, Russia\\~\\
	$^6$Swiss-Norwegian Beamlines at the European Synchrotron Radiation Facility, 38000 Grenoble, France\\~\\
	$^7$Saint-Petersburg State University, Ulyanovskaya 1, Saint-Petersburg, 198504, Russia\\~\\
	$^8$Kapitza Institute for Physical Problems of RAS, Kosygina street, 2, Moscow, 119334, Russia\\~\\
}

\dates{18 May 2016}{today}

\abstract{
	Magnetic properties of single crystals of Mn$_{1-x}$Fe$_x$Si solid
	solutions with $x < 0.2$ are investigated by pulsed field technique in
	magnetic fields up to 50~T.  It is shown that magnetization of Mn$_{1-x}$Fe$_x$Si in
	the paramagnetic phase follows power law $M(B) \sim B^\alpha$ with the
	exponents $\alpha \sim 0.33-0.5$, which starts above characteristic
	fields $B_c \sim 1.5-7$~T depending on the sample composition and lasts
	up to highest used magnetic field. Analysis of magnetization data
	including SQUID measurements in magnetic fields below 5~T suggests that
	this anomalous behavior may be likely attributed to the formation of a
	field-induced Griffiths phase in the presence of spin-polaron effects.
}


\begin{document}

\maketitle

1. Substitutional solid solutions manganese monosilicide --- iron monosilicide,
Mn$_{1-x}$Fe$_x$Si attract attention due to unique set of their physical properties, which
include quantum critical phenomena \cite{Bauer10,Demishev13},
spiral magnetic order \cite{Grigoriev09, Grigoriev10, Grigoriev11} and
development of the magnetic phases with short-range magnetic order
\cite{Bauer10,Demishev13,Bauer12,Hamann11,Tewari06,Kruger12,Demishev16},
similar to blue fog phases in liquid crystals
\cite{Bauer12,Hamann11,Tewari06,Kruger12} or to spin-liquid phases
\cite{Demishev16}.  At the same time, the essential fundamental questions concerning the
nature of magnetism in Mn$_{1-x}$Fe$_x$Si system remain unsolved. It was taken for granted
during decades that magnetic properties of MnSi, FeSi and MnSi based solids may
be adequately described with the help of an itinerant model, which assumes a
crucial role of spin fluctuations together with distributed spin density in the
unit cell \cite{Moriya85}. This point of view contradicts to recent electron spin resonance
(ESR) experiments demonstrating localized character of magnetic moments in
Mn$_{1-x}$Fe$_x$Si \cite{Demishev11,Demishev14} and to observation of
the Yosida-type magnetic scattering \cite{Yosida57}
on localized magnetic moments, which dominates in magnetotransport properties
\cite{Demishev16,Demishev12}. Theoretical calculations in the framework of local density
approximation (LDA) technique show that spin density in MnSi is localized on Mn
ions, rather than being smeared \cite{Corti07}. It is worth noting that, in the
fundamental work by Maleyev \cite{Maleyev06} as well as in the
subsequent publications \cite{Grigoriev09,Grigoriev10,Grigoriev11},
the successful theoretical accounting of the magnetic properties of Mn$_{1-x}$Fe$_x$Si is
\emph{de-facto} based on Heisenberg model of magnetism, i.e. implies presence of
localized magnetic moments (LMM).

However, for resolving the paradigm of LMM-based magnetism in Mn$_{1-x}$Fe$_x$Si, it is
necessary to explain the reduced value of saturated magnetization (less than
Bohr magneton $\mu_B$ per Mn ion) when manganese LMM is about $\mu_{\mathrm{Mn}}
\sim 1.2 \mu_B$ \cite{Corti07} and to suggest consistent explanation of ESR and magnetic
scattering experiments \cite{Demishev11,Demishev14,Demishev12} together with specific features of the field
and temperature dependences of magnetization $M(B, T)$ \cite{Demishev15}. For this purpose, a
spin polaron phenomenological model was developed, where spin polaron represents
a nanometer size quasi-bound state of itinerant electrons in the vicinity of
manganese localized magnetic moments \cite{Demishev11,Demishev12,Demishev15}.  Opposite orientation of Mn LMM
and magnetic moments of band electrons results in reduction of the saturated
magnetization and transitions of electrons between the quasi-bound and band
states thus leading to spin fluctuations. This model construction is not
confirmed by direct structural studies of the nanoscale so far, but spin polaron
hypothesis has been successfully applied to the explanation of experimental data
\cite{Demishev11,Demishev14,Demishev12,Demishev15}, which contradict to a widely accepted model of itinerant
magnetism \cite{Moriya85}. Moreover, as long as spin polaron represents an elementary
ferrimagnet, the considered model opens an opportunity for natural
interpretation of the recently discovered striking similarity between physical
properties of metallic MnSi and ferrimagnetic dielectric Cu$_2$OSeO$_3$ with
Heisenberg LMM \cite{Sidorov14}, which cannot be either foreseen or explained in an
itinerant model.

In order to make the right choice between the competing models of magnetism of
Mn$_{1-x}$Fe$_x$Si, it is instructive to study magnetic properties in high magnetic
fields. Indeed magnetic field may affect both spin fluctuations and spins
alignment in spin polaron, so that analysis of the $M(B, T)$ data may shed more
light on the origin of the magnetism in this system. At present there is a
limited set of high field magnetization data for MnSi only, and field
dependences $M(B, T = \mathrm{const})$ are reported up to 50~T for the magnetically ordered
spiral phase, and up to 30~T in the paramagnetic phase \cite{Sakakibara82}. To the best of our
knowledge, the magnetization of substitutional solid solutions Mn$_{1-x}$Fe$_x$Si has
never been examined in strong magnetic field. In the present work, we undertake
the investigation of the magnetization in the paramagnetic phase of Mn$_{1-x}$Fe$_x$Si
with $x < 0.2$ in pulsed magnetic fields up to 50~T.

2. Single crystals of Mn$_{1-x}$Fe$_x$Si with $x < 0.2$ investigated in the present work
were identical to those studied in Ref.~\citen{Demishev13}. The quality of the samples was
controlled by X-ray and neutron diffraction. Samples composition was determined
by electron probe micro-analysis (EPMA). The deviation from stoichiometric
composition between metals (Fe, Mn) and silicon 1:1 did not exceed the value
$\sim 0.5\%$ comparable with absolute error of our EPMA measurements. Field and
temperature dependences of the magnetization in the magnetic field below 5~T
were measured by using SQUID magnetometer (Quantum Design). Experiments in
magnetic fields up to 50~T were conducted at KU Leuven pulsed field facility
\cite{Vanacken13} (Belgium). In all cases, the magnetic field was aligned along [110]
direction. The temperatures, corresponding to the paramagnetic (PM) phase of
Mn$_{1-x}$Fe$_x$Si may be defined as $T > T_s(x)$, where $T_s(x)$ marks the transition into
magnetic phase with short-range magnetic order \cite{Demishev13,Demishev16}. The values of $T_s(x)$ were
chosen in accordance with $T$-$x$ magnetic phase diagram obtained in \cite{Demishev13}.

Pulsed field measurements of the magnetization field dependences were carried
out with the help of the installation based on induction technique \cite{Lagutin95}. In our
case, the difference with respect to Ref.~\citen{Lagutin95} consisted in that the needle-like
sample was located inside compensated pick-up coils. It is known that
magnetization measurements of pure MnSi with pulsed magnetic field are affected
by heating effects, which lead to complicated procedure of correct experimental
differential of the $M(B)$ curve \cite{Sakakibara82}. Heating of the sample may depend on
thermodynamic contribution to free energy, which scales with the sample
magnetization \cite{Sakakibara82}, and may be caused by Joule heating by induction currents in
metallic Mn$_{1-x}$Fe$_x$Si samples. For that reason, in order to reduce possible
temperature variation, the samples of iron concentration $x = 0.054$,
$x = 0.11$ and $x = 0.19$ were chosen for pulsed field experiments. In this
concentration range magnetization of Mn$_{1-x}$Fe$_x$Si is at least two times less
than that of pure MnSi \cite{Demishev13,Demishev12} and resistivity is about an order of magnitude
higher.

% 2016-06-06 {
For these samples the transition temperatures into the phase with short-range magnetic order
are $T_s(x = 0.054) = 17.8$~K, $T_s(x = 0.11) = 8.9$~K and $T_s(x = 0.19) = 3.1$~K \cite{Demishev13}.
The sample with iron concentration $x = 0.054$ possesses Curie temperature $T_c = 12.6$~K, which is
about 2.3 times less than in pure MnSi. At the same time, the transition into the phase with
long-range magnetic order transition is strongly suppressed for $x \ge 0.11$ and does not
exceed $\sim 0.6$~K \cite{Demishev13,Demishev16}.
% }

As long as in induction pulsed field experiments both magnetization $M(t)$
and field $B(t)$ are functions of time $t$, change of the sample temperature will not be
uniform during the pulse depending on the thermal exchange and thermal diffusion
times. Careful comparative analysis of the $M(B)$ curves corresponding to pulses
$B(t)$ with different magnitude allowed to conclude that the part of $M(t)$ pulse
corresponding to increasing magnetic field was not affected by temperature
variation in the studied samples. Therefore below we will consider only this
part of pulsed field measurements, where the condition $T(t) = \mathrm{const}$
is valid. Additional argument favoring the above assumption is the excellent
reproduction of the $M(B)$ curves shape in pulsed field and steady field
measurements performed at the same temperature. The latter observation also
allowed using the magnetization curves obtained with the help of SQUID
magnetometer for the absolute calibration of pulsed field data.

\begin{figure*}[t]
	{\center
		\includegraphics[width=1\textwidth]{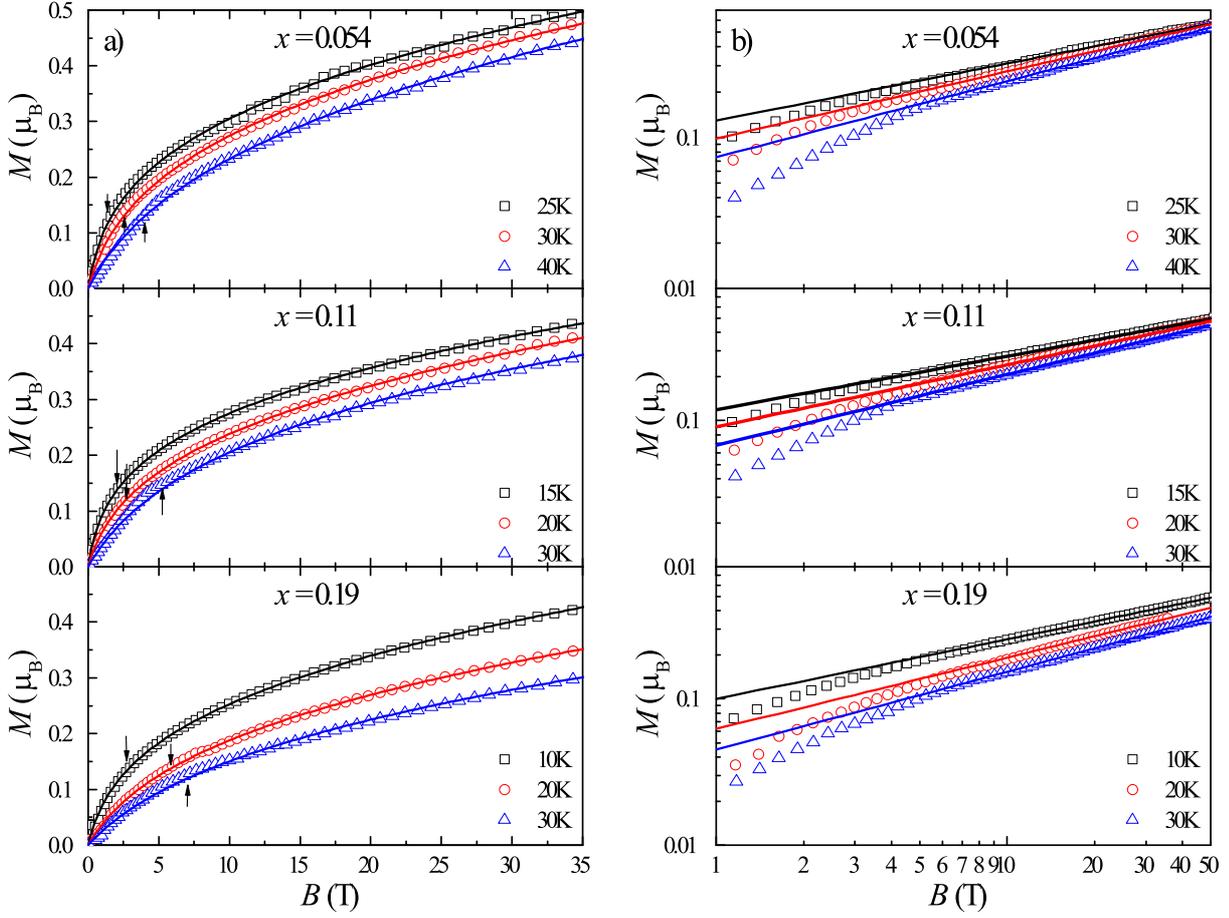}
	}
	%\missingfigure[figwidth=\textwidth]{}
	%\missingfigure[figwidth=\textwidth, figheight=5cm]{Figure 1}
	\caption{
		\label{fig1}
		Fig.~\ref{fig1}.
		Magnetization field dependences for Mn$_{1-x}$Fe$_x$Si.
		Points in the panels a, b --- experiment,
		lines in the panel b --- best fit by the power law
		(Equation~\ref{eq1}), lines in the panel a --- best fit
		with the help of interpolating formula (Equation~\ref{eq3}).
		Arrows in the panel a denote crossover fields $B_c$.
	}
\end{figure*}

3. We first analyze $M(B)$ data in high magnetic fields. It is remarkable that up
to the highest magnetic fields studied, field dependences of the magnetization
in the studied samples do not saturate (Fig.~\ref{fig1}, panels a-b). Interesting that
double logarithmic plot (Fig.~\ref{fig1},b) suggests high-field power behavior

\begin{equation} \label{eq1}
	M(B, T) = A(T) \cdot B^{\alpha(T)},
\end{equation}

\noindent where both pre-factor $A$ and exponent $\alpha < 1$ depend on
temperature and concentration. This observation is very unusual, because since
the pioneering work on MnSi \cite{Sakakibara82} it is believed that magnetization in strong
magnetic field increases linearly with constant slope, and therefore it is
natural to expect similar behavior in Mn$_{1-x}$Fe$_x$Si at least for the samples with
low iron concentration.

In order to analyze high-field asymptotics of the magnetization field dependence
in more detail, it is instructive to consider the function
$F(B) = \partial{M}/\partial{\mathrm{ln}B} - M(B)$, which may be
computed for each measured $M(B)$ curve. If Equation (\ref{eq1}) is valid,
the $F(B)$ is given by

\begin{equation} \label{eq2}
	F(B) = \frac{\partial{M}}{\partial{\mathrm{ln}B}} - M(B) =
		A(T) \cdot (\alpha(T) - 1) \cdot B^{\alpha(T)},
\end{equation}

\noindent and hence in high magnetic field the $F(B)$ field dependence will reproduce the
same power law except the case $\alpha = 1$, for which $F(B) \equiv 0$. In the case of a
paramagnet with saturating magnetization $M(B \to \infty) = M_0$, the function $F(B)$ will also
saturate at the value $F(B \to \infty) = -M_0$, because for $B \to \infty$ the
derivative $\partial{M}/\partial{\mathrm{ln}B}$ turns to zero. Consequently the
behavior of $F(B)$ allows discriminating between different cases,
which may be expected for Mn$_{1-x}$Fe$_x$Si.

\begin{figure}[t]
	\includegraphics[width=1\linewidth]{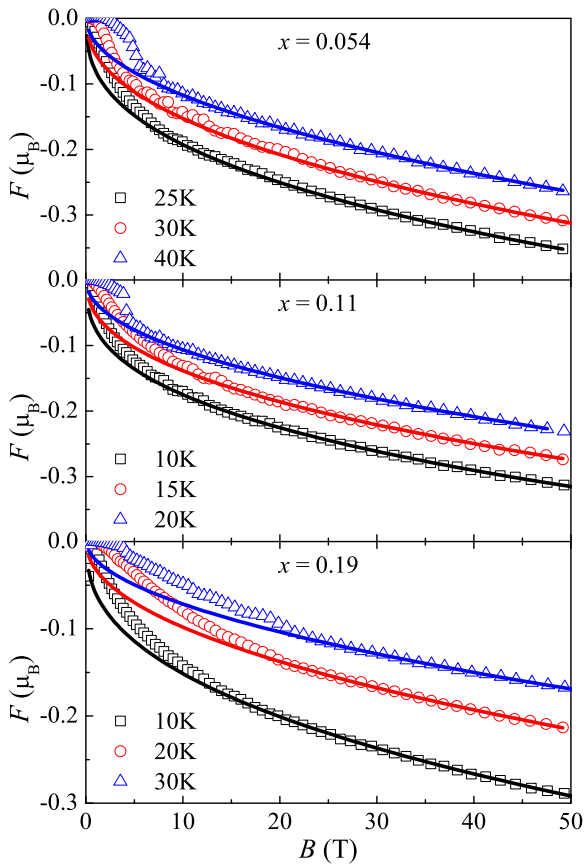}
	%\missingfigure{Figure 2}
	\caption{
		\label{fig2}
		Fig.~\ref{fig2}.
		Function $F(B)$.
		Points --- calculation from experimental data
		in Fig.~\ref{fig1}; lines --- approximation by
		Equation (\ref{eq2}).
		Parameters of the approximations are the same
		as in Fig.~\ref{fig1},b.
	}
\end{figure}

\begin{figure}[t]
	\includegraphics[width=1\linewidth]{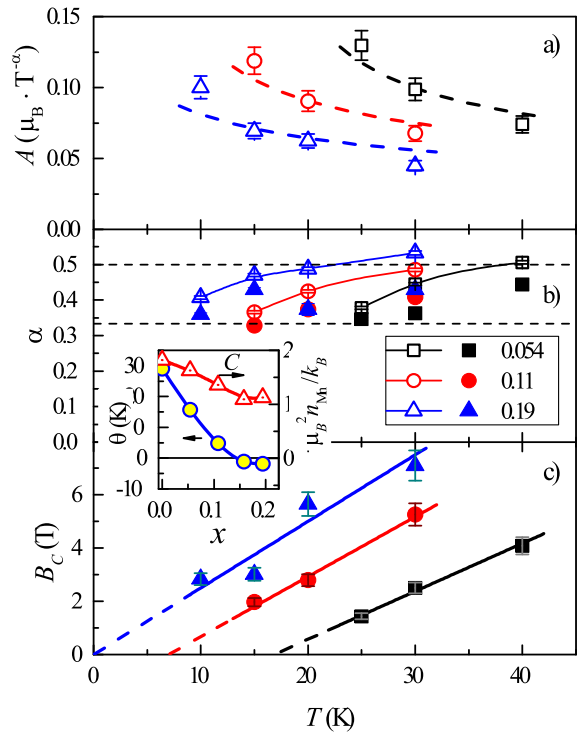}
	%\missingfigure{Figure 3}
	\caption{
		\label{fig3}
		Fig.~\ref{fig3}.
		Parameters in the model dependences: coefficient $A$ in
		Equation~\ref{eq1} (panel~a), the exponents $\alpha$ (panel~b)
		and crossover fields $B_c$ (panel~c).
		Open symbols denote results obtained from fitting by
		Equation~\ref{eq1}; solid symbols correspond to approximation
		with the help of interpolation Equation~\ref{eq3}.
		Inset shows concentration dependences of Curie constant $C$ and
		paramagnetic temperature $\theta$ obtained from the magnetic
		measurements in a low magnetic field.
	}
\end{figure}

Calculated functions $F(B)$ are presented in Fig.~\ref{fig2}. It is visible, that the
suggested in \cite{Sakakibara82} form for high-field asymptotics
$M(B) = M_0 + \chi^{*} \cdot B$, where $\chi^{*}$ denotes some effective
susceptibility, does not meet experiment, as long as in this case
$F(B)$ should saturate at some negative value. Instead of this, experimental data
can be well fitted with the help of Equation (\ref{eq2}) (solid lines in
Fig.~\ref{fig2}) with $A$ and $\alpha$ as free parameters.
The corresponding approximations by power law (\ref{eq1}) with the same $A$ and
$\alpha$ are shown in Fig.~\ref{fig1},b (solid lines).
The parameter $A(T)$ decreases with temperature and iron concentration
(Fig.~\ref{fig3},b), whereas the exponent of the power law (\ref{eq1})
demonstrates opposite behavior: $\alpha(T)$ increases with $T$ and $x$
and reaches value $\alpha(T) \sim 0.5$ at $T \sim 20-40$~K
(open symbols in Fig.~\ref{fig3},b).

Now we will examine the region of a weak magnetic field. Straightforward
analysis \cite{Supplemental} shows that in the paramagnetic phase of
Mn$_{1-x}$Fe$_x$Si the Curie-Weiss asymptotics $M(B) = C \cdot B / (T - \theta)$ hold
for temperatures which are higher by 1-3~K than $T_s(x)$ \cite{Demishev16}.
% 2016-06-06 {
Hereafter $C$ is Curie constant and parameter $\theta$ is denoted as the paramagnetic
temperature. The parameters $C(x)$ and $\theta(x)$ decrease with $x$ and
the paramagnetic temperature changes sign at $x \sim 0.12-0.15$ (inset in Fig.~\ref{fig3}).
% }
It is worth noting that the change of the sign of the
Mn-Mn exchange energy $J$ from ferromagnetic ($J > 0$)
to antiferromagnetic ($J < 0$) was recently predicted for Mn$_{1-x}$Fe$_x$Si with
$x \sim 0.17$ \cite{Glushkov15}. The obtained dependence $\theta(x)$ reasonably concurs with this result.

In order to describe field-induced transformation of the Curie-Weiss linear
dependence $M(B) \sim B$ into the power law $M(B) \sim B^\alpha$ the following interpolating
formula may be used

\begin{equation} \label{eq3}
	M(B) = \frac{C}{T - \theta} \cdot \frac{B}{
		[B / B_c(T) + 1]^{1 - \alpha}
	},
\end{equation}

\noindent where $B_c(T)$ denotes the crossover field. Equation (\ref{eq3}) gives
$M(B) = C \cdot B/(T - \theta)$ for $B \ll B_c(T)$ and corresponds to the power
dependence (\ref{eq1}) with $A(T) = C \cdot B_c^{1 - \alpha} / (T - \theta)$
for $B \gg B_c(T)$. As long as parameters $C$ and $\theta$ are known \cite{Supplemental}, the interpolating
formula (\ref{eq3}) provides two-parameter ($B_c$ and $\alpha$) approximation for a magnetization
field dependence at fixed temperature. The results of fitting with the help of
Equation (\ref{eq3}) are presented in Fig.~\ref{fig1},a by solid lines. It is possible to
conclude that this approach describes experimental data reasonably well. The
obtained values of $B_c(T)$ are shown in Fig.~\ref{fig3},c and marked by arrows in
Fig.~\ref{fig1},a. The crossover field increases with temperature almost linearly and
extrapolation of the $B_c(T)$ dependence to the value $B_c = 0$ gives characteristic
temperatures, which are very close to $\theta(x)$ (inset in Fig.~\ref{fig3}).
Therefore in the studied system $B_c(T)$ approximately follows the law
$B_c \sim (T - \theta)$ with almost the same paramagnetic temperature as in
Curie-Weiss asymptotics. In addition, the
analysis of the experimental data by Equation (\ref{eq3}) gives exponents
$\alpha(T)$ which are somewhat different from those found from the power law approximation (see
solid and open symbols in Fig.~\ref{fig3},b respectively). In the considered case, the
average value is $\langle\alpha\rangle = 0.38 \pm 0.04$. Therefore it is possible to get an approximate
expression for the coefficient $A$ in the power law (\ref{eq1}),
$A(T) = C \cdot B_c^{1 - \alpha} / (T - \theta) \approx
A_0 / (T - \theta)^{\langle\alpha\rangle}$, where $A_0$ is a numerical coefficient depending
on $C$ and $dB_c / dT$.
Dashed line in Fig.~\ref{fig3},a are drawn in accordance with this estimate and it can be
seen that deduced $A(T)$ function reasonably reproduces results of fitting by
power dependence (\ref{eq1}) without introducing any new free parameters. This
consideration shows that the low magnetic field asymptotics and high magnetic
field asymptotics of $M(B)$ are closely linked in Mn$_{1-x}$Fe$_x$Si, and Curie-Weiss
behavior transforms into a power law when magnetic field increases.

4. The power asymptotics $M(B) \sim B^\alpha$, which are observed instead of saturated
magnetization, are very unusual for several reasons. Firstly, existing theories
predict the power law for magnetization when the system ground state is a
Griffiths phase \cite{Griffiths69,Bray87,Demishev10,Dasgupta85,Fisher92,Fisher95}.
This behavior may be expected in magnetically
disordered systems in the case $k_B T \ll \mu^{*} B$ with either ferromagnetic
\cite{Fisher92,Fisher95}, or
antiferromagnetic \cite{Dasgupta85} interactions (here $\mu^{*}$ denotes the effective magnetic
moment). The calculated values of the exponent $\alpha$ lie within limits
$0.2 \leq \alpha \leq 0.6$ \cite{Dasgupta85} or $1/3 \leq \alpha \leq 1/2$ \cite{Fisher95} in agreement
with the experimental data (Fig.~\ref{fig3},b).
However, the aforementioned theoretical results were obtained for chain systems
\cite{Dasgupta85,Fisher92,Fisher95}. Moreover, in the early work \cite{Bulaevskii72} the field-induced crossover between
$M(B) \sim B$ and $M(B) \sim B^\alpha$ was discovered in TCNQ based conducting spin chains, and was
explained theoretically as a characteristic feature of just one-dimensional case
without any disorder.

Apparently, Mn$_{1-x}$Fe$_x$Si is a three-dimensional system, and results of Ref.~\citen{Bulaevskii72} are
not applicable to our case. At the same time, Griffiths phase scenario is more
general \cite{Demishev10,Dasgupta85}, and in our opinion may be considered as a hypothesis to explain
our results. Magnetic system in the Griffiths phase becomes separated into spin
clusters due to spatial dispersion of exchange energies. The spin clusters with
a higher (as compared to the average value) degree of correlation determine
\cite{Bray87,Demishev10,Dasgupta85,Fisher92,Fisher95} the Griffiths phase magnetic properties. On a qualitative level,
variations in temperature or magnetic field result in "scanning" over the
randomly distributed spin clusters parameters \cite{Bray87,Demishev10} and this could lead to a
departure from Curie-Weiss law for magnetic susceptibility $\chi(T)$ at low
temperatures ($\chi(T)$ for $T \to 0$ diverges as a power of inversed temperature
\cite{Bray87,Demishev10,Dasgupta85,Fisher92,Fisher95})
and to asymptotics $M(B) \sim B^\alpha$ \cite{Dasgupta85,Fisher92,Fisher95}.
However, in the studied case, no deviations
from Curie-Weiss law in weak magnetic field are observed \cite{Supplemental}. This means that
if the concept of a Griffiths phase is applied to Mn$_{1-x}$Fe$_x$Si samples studied in
the present work, it is necessary to assume that Griffiths phase forms a high
magnetic field $B > B_c$ and is missing in a weak magnetic field $B < B_c$.

Anyhow, the implementation of the Griffiths phase concept demands onset of
magnetic inhomogeneity. This constitutes the second problem, as long as
Mn$_{1-x}$Fe$_x$Si is known as a system with strong spin fluctuations \cite{Moriya85,Demishev14}. Spin
density in itinerant magnetism model is distributed in the unit cell and this
distribution is kept by spin fluctuations \cite{Moriya85}. Therefore there is no way to
obtain spatial magnetic inhomogeneity even in the case, when spin fluctuation
rate is reduced in an external magnetic field. Therefore, the experimental
results obtained in the present work do not favor itinerant description of
magnetic properties of Mn$_{1-x}$Fe$_x$Si.

At the same time the application of the phenomenological spin-polaron model \cite{Demishev15}
is not straightforward. From one hand, as long as spin polarons are kind of
magnetic inhomogeneities on the nanoscale, spin polaron paradigm looks promising
for interpretation based on the Griffiths phase formation. According to \cite{Demishev15}, in
the case when the concentration of electrons per Mn $n_e / n_{\mathrm{Mn}}$ satisfies inequality
$n_e / n_{\mathrm{Mn}} < (\mu_{\mathrm{Mn}}/\mu_e)^2$ the magnetic subsystem of Mn$_{1-x}$Fe$_x$Si will demonstrate magnetic
phase separation into two types of magnetic states: spin polarons and bare Mn
ions (here $\mu_{\mathrm{Mn}}$ and $\mu_e$ denotes magnetic moments of Mn LMM and band electron
respectively). Repeating calculations of Ref.~\cite{Demishev15}, it is possible to show that for
any ratio of the numbers of spin-polaron states and bare Mn ions, Curie constant
acquires the form $C = n_{\mathrm{Mn}} \mu_{\mathrm{Mn}}^2 / k_B$, and therefore experimental data (inset in
Fig.~\ref{fig3}) correspond to the Mn localized magnetic moments
$\mu_{\mathrm{Mn}} = (1.1-1.3)\mu_B$. This
estimate correlate well with the LDA calculations, which give
$\mu_{\mathrm{Mn}} = 1.2 \mu_B$. Hall effect measurements give $n_e / n_{\mathrm{Mn}} = 0.9$ \cite{Glushkov15},
so that condition for magnetic inhomogeneity is fulfilled for all Mn$_{1-x}$Fe$_x$Si samples studied.

However, it is necessary to explain how possible magnetic inhomogeneity
transforms in a strong magnetic field into disordered spin configuration
specific to a Griffiths phase. This is hardly possible without elucidation of
the microscopic nature of the proposed phenomenological spin-polaron model \cite{Demishev15},
which will be a subject of future investigations (both theoretical and
experimental). Here we would like to make only several general remarks. First
remark is connected with the correspondence between the spin-polaron state
introduced for Mn$_{1-x}$Fe$_x$Si \cite{Demishev15} and the standard Nagaev-Mott-Kasuya-Krivoglaz
\cite{Mott71,Nagaev68,Kasuya70,Krivoglaz74} ferromagnetic (FM) polarons or ferrons, which appear in the
antiferromagnetic (AFM) or PM matrices in the double-exchange model
of De~Gennes \cite{Gennes60} or in the FM Kondo-lattice model (FM KLM). These models are
relevant for manganites and other systems exhibiting colossal magnetoresistance
phenomenon. It is possible to note many similarities between ferrons in the
double exchange model and considered spin-polarons with respect to the
percolative nature of the phase-transitions to FM state \cite{Kagan01}, the general
expression for Curie-Weiss magnetic susceptibility in phase-separated state \cite{Kagan13}
and so on. At the same time, there is a striking difference in high magnetic
field behavior, where ferrons are growing both at $T = 0$ and at finite temperatures
\cite{Sboychakov02}, which leads to saturation of the magnetization instead of the power-law
dependence in the studied case. Note that for ferrons description in the FM KLM
we should have a FM sign ($J > 0$) of the exchange (Hund’s) integral $J$ between
local spins and spins of conductivity electrons which are parallel for large $J$.
As long as the sign of this exchange in Mn$_{1-x}$Fe$_x$Si is not known exactly, and
interaction of AFM nature cannot be excluded \emph{a priori} (for example, in pure MnSi
we should rather expect $J > 0$ in analogy with manganites, while in pure FeSi
probably $J < 0$), then instead of FM KLM it will be necessary to consider an AFM
KLM \cite{Kondo64,Tsvelik98,Fulde93} with quite different physics. Namely, with large absolute values of
$J$ we may have the local singlets or generalized Kondo-singlets (totally screened
or overscreened as in multi-channel KLM \cite{Tsvelik98}). Note that in cuprates the local
singlets on elementary CuO$_4$ plaquette are just famous Zhang-Rice singlets, which
reduce the two-band Emery model to the one band $t$-$J$ model \cite{Zhang88}. When extending
this analogy to the case of Mn$_{1-x}$Fe$_x$Si, we can possibly speak with some
precautions about the totally screened (with respect to spin) or partially
screened (ferrimagnetic) complexes consisting of 1 Mn ion and 1 or 2
conductivity electrons (or of even 2 neighbouring Mn ions and 2 or 3
conductivity electrons quasilocalized on these ions). It is worth noting that a
complex structure of spin-polaron consisting of several Mn ions and electrons
was introduced in \cite{Demishev15}. Moreover, the application of the thermodynamic stability
requirement \cite{Demishev15} will result in allmost antiparallel configuration of local
spins and spins of conductivity electrons in small magnetic fields. This
picture, however, is an oversimplification again since it totally neglects the
charge degrees of freedom. The more general situation corresponds to the
periodic Anderson model \cite{Anderson70} with strong hybridization of p-orbitals of Si and
d-orbitals of transition element. Interesting, that according to Ref.~\citen{Carbone06} an
intermediate mixed-valence state is expected for pure MnSi \cite{Carbone06} and it is
possible that the same situation holds in Mn$_{1-x}$Fe$_x$Si for moderate Fe
concentrations. This means that in correct microscopic model both spin and
charge sectors should be taken into account on equal grounds. In particular, in
charge sector an electron-polaron effect (connected with the interband  Hubbard
interaction) is very important \cite{Nozieres69,Kagan87,Kagan11}. In spin sector we should also consider
the competition between the spin fluctuations and RKKY-interaction of Mn LMM
expected for Mn$_{1-x}$Fe$_x$Si \cite{Glushkov15}, which may lead to a spin-glass phase (or a
Griffiths phase) in agreement with generalized Mott-Doniach mechanism \cite{Mott74,Doniach67}. 

For substantial Fe concentrations it is possible to suppose that in high
magnetic field $B > B_c(T)$ new magnetic phase, characterized by power law
$M(B) \sim B^\alpha$
develops. In this case, the crossover field $B_c$ may be linked with the spin
fluctuations magnitude. For $B < B_c$ these fluctuations are strong enough and system
is in paramagnetic phase, whereas for $B > B_c$ field-induced weakening of spin
fluctuations gives way to a Griffiths-type magnetic phase. Moreover, it is
natural to expect that crossover field may increase with temperature as long as
the magnitude of spin fluctuations will increase with temperature in both
itinerant \cite{Moriya85} and spin-polaron models \cite{Demishev14,Demishev15}. In addition, the increase of the
spin fluctuations magnitude with iron concentration discovered in \cite{Demishev14} should
result in a corresponding enhancement of $B_c$. These qualitative speculations
completely meet experiment, where $B_c(T, x)$ increases with both $T$ and $x$
(Fig.~\ref{fig3},c). 

\begin{figure}[t]
	\includegraphics[width=1\linewidth]{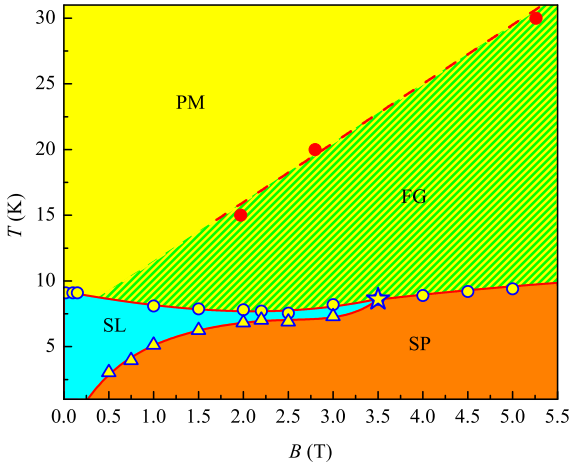}
	%\missingfigure{Figure 4}
	\caption{
		\label{fig4}
		Fig.~\ref{fig4}.
		Magnetic phase diagram for the sample with $x = 0.11$.
		Open symbols --- data from Ref.~10,
		solid symbols --- crossover field $B_c(T)$,
		which mark expected transition into field-induced Griffiths
		phase (FG). Different magnetic phases in accordance with
		Ref.~10 are paramagnetic phase (PM), spin-liquid phase
		(SL) and spin-polarized phase (SP).
	}
\end{figure}

It is interesting to compare results of the present investigation with the phase
diagram in a magnetic field recently obtained for the sample with $x = 0.11$ \cite{Demishev16}.
It was found that the magnetic phase diagram is formed by paramagnetic (PM),
spin-liquid (SL) and spin-polarized (SP) phases (Fig.~\ref{fig4}). This work adds new
crossover line inside PM phase $B_c(T)$, which separate different regimes of spin
fluctuations and marks onset of the field-induced Griffiths phase (FG)
consisting of spin clusters. Successful application of the spin-polaron model
for the explanation of the peculiarities in the paramagnetic phase strongly
supports supposition that spin polarons may be considered as a "building blocks"
for various magnetic phases in the case of Mn$_{1-x}$Fe$_x$Si (Fig.~\ref{fig4}). Within this
concept we expect that the singular point discovered in \cite{Demishev16} (star in
Fig.~\ref{fig4}) should be a triple point rather than a critical point as expected in
spin-polaron scenario \cite{Demishev16}. In strong magnetic field $B > 3.5$~T the increase in
temperature at first leads to melting of the spin-polarized phase characterized
by parallel spin alignment into special spin-cluster based phase (FG). Further
increase of temperature will result in a crossover-type transition from FG with
suppressed spin fluctuations to an effectively homogeneous PM phase with strong
spin fluctuations (Fig.~\ref{fig4}). More complicated behavior may be expected for
intermediate fields $B \sim 1$~T, where variation of temperature may induce a sequence
of phase transitions between the spin-polarized, spin-liquid, field-induced
Griffiths and paramagnetic phases (Fig.~\ref{fig4}). These phases may demonstrate
different correlation lengths and regimes of spin fluctuations, so that physical
picture in a magnetic field should go much further than often considered a
simple fluctuation scenario for $B = 0$ \cite{Grigoriev11,Janoschek13}. For that reason we expect that
additional studies of magnetic structure and magnetic fluctuations with the help
of advanced neutron scattering technique could be rewarding and may shed more
light on the nature of spin-polaron states and complicated magnetic phase
diagram of Mn$_{1-x}$Fe$_x$Si solid solutions.

5. In conclusion, we have shown that the magnetization of Mn$_{1-x}$Fe$_x$Si in the
paramagnetic phase follows power law $M(B) \sim B^\alpha$ with the exponents
$\alpha \sim 0.33-0.5$, which starts above characteristic fields
$B_c \sim 1.5-7$~T depending on the sample composition and lasts up to 50~T.
In contrast to previous experimental and theoretical studies, the asymptotic
$M(B) \sim B^\alpha$ behavior is observed in three-dimensional case rather
than in one-dimensional spin chain system. This
anomalous behavior may be attributed to the possible formation of a
field-induced Griffiths phase presumably caused by spin-polaron effects. 

Authors are grateful to D.I.~Khomskii for helpful discussions. This work was
supported by programmes of Russian Academy of Sciences "Electron spin resonance,
spin-dependent electronic effects and spin technologies", "Electron correlations
in strongly interacting systems". The work at KU Leuven was supported by the
Methusalem Funding by the Flemish Government. M.Yu.~K. thanks the Program of Basic
Research of the National Research University Higher School of Economics for
support.

%\bibliography{MnFeSi_1.bib}
%\bibliographystyle{plain}

%\listoftodos

\end{document}

\maketitle

\begin{figure*}[h]
	\centering
	\includegraphics[width=0.6\textwidth]{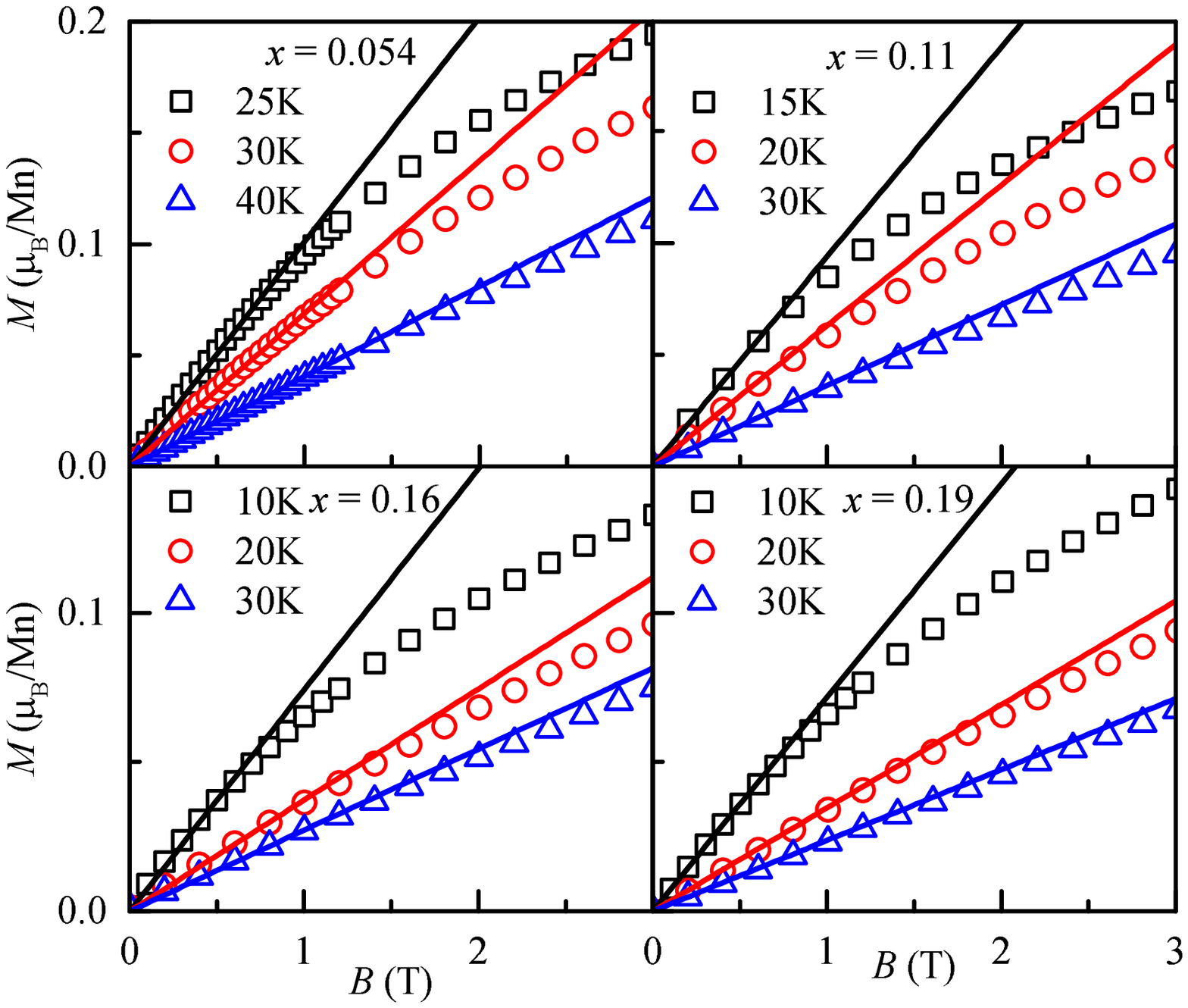}
	\caption{
		\label{fig:supp1}
		Fig.~S\ref{fig:supp1}.
		Typical magnetization field dependences for different
		Mn$_{1-x}$Fe$_x$Si samples in a weak magnetic field.
		Points --- experiment, lines --- approximation by linear
		law $M(B, T) = \chi_0(T) \cdot B$.
	}
\end{figure*}

\begin{figure*}[h]
	\centering
	\includegraphics[width=0.6\textwidth]{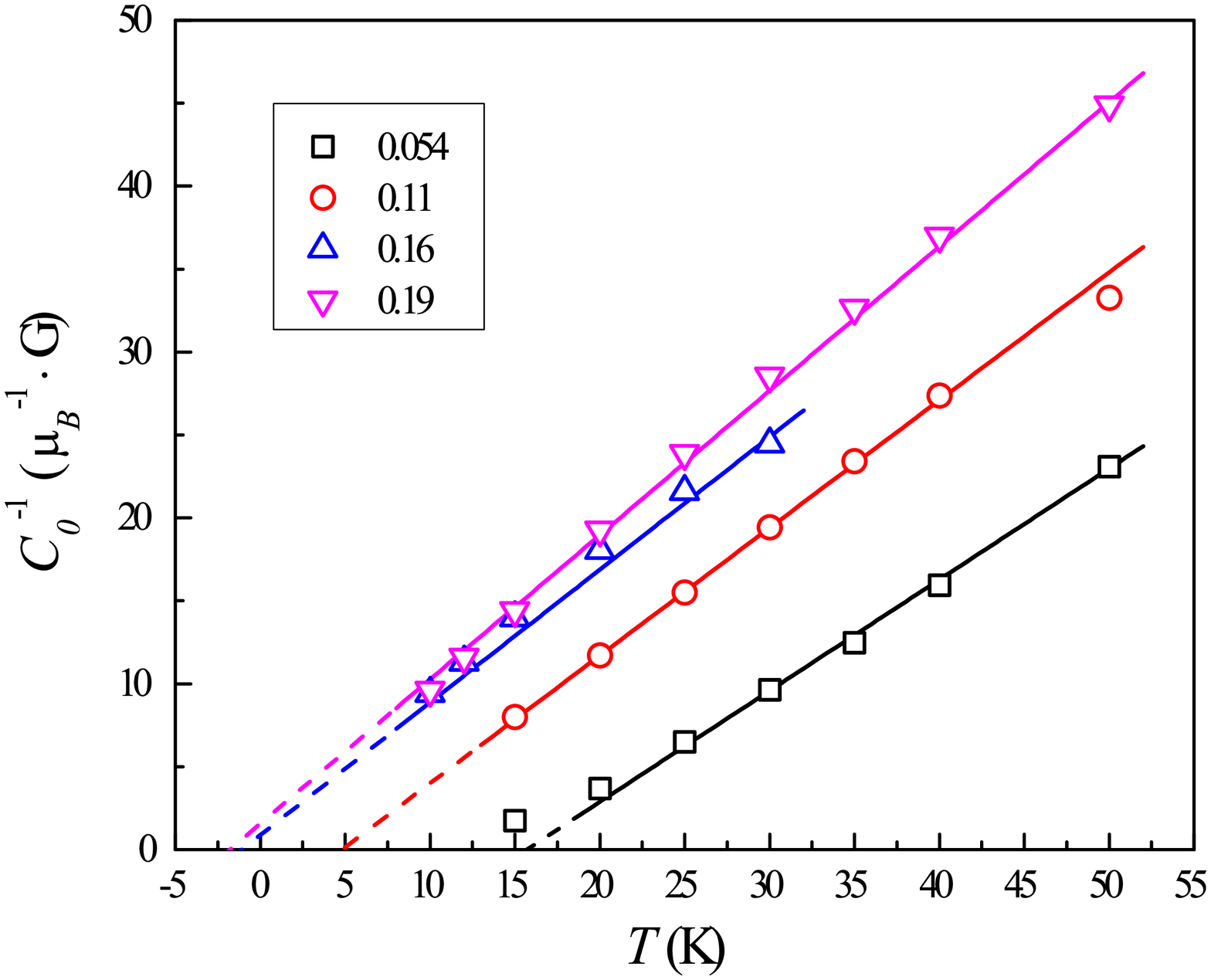}
	\caption{
		\label{fig:supp2}
		Fig.~S\ref{fig:supp2}.
		Temperature dependences of the susceptibility $\chi_0(T)$
		in coordinates $\chi_0^{-1} = f(T)$. Points --- data
		obtained from analysis of the $M(B, T = \mathrm{const})$
		curves, lines --- best fits with the help of
		equation $\chi_0^{-1} = (T - \theta) / C$.
	}
\end{figure*}

Field dependences of magnetization at fixed temperatures
$M(B, T = \mathrm{const})$ were measured with the help of SQUID magnetometer
(Quantum Design) up to $B = 5$~T for the Mn$_{1-x}$Fe$_x$Si samples with
iron concentration $x < 0.2$. Typical experimental results are shown in
Fig.~S\ref{fig:supp1}. Low field sections of $M(B, T = \mathrm{const})$
curves were approximated by linear law $M(B, T) = \chi_0(T) \cdot B$ (lines
in Fig.~S\ref{fig:supp1}). The obtained $\chi_0(T)$ dependences for each
particular sample were analyzed in coordinates $\chi_0^{-1} = f(T)$
(Fig.~S\ref{fig:supp2}). The obtained $\chi_0^{-1} = f(T)$ dependences are
linear and may be presented in the form $\chi_0^{-1} = (T - \theta) / C$
corresponding to Curie-Weiss law with Curie constant $C$ and paramagnetic
temperature $\theta$. These parameters were found from linear fits of the
data in Fig.~S\ref{fig:supp2} and shown at the inset in Fig.~3 of
the main text.

\newpage\clearpage

